\documentclass[12pt]{article}
{\baselineskip 12pt}
\begin{document}
\title{
\vspace*{-30pt}
On masses of unstable particles \\ and their
antiparticles  \\in the CPT--invariant system.}
\author{ \hfill \\ K. Urbanowski$^{\ast}$ \\  \hfill  \\
University of Zielona Gora, Institute of Physics, \\
ul. Podgorna 50, 65-246 Zielona Gora, Poland. \vspace*{-10pt}}
\maketitle {\noindent}{\em PACS numbers:}  03.65.Ca., 11.30.Er.,
11.10.St., 14.40.Aq. 
\begin{abstract}
We show that the diagonal matrix elements of the effective
Hamiltonian governing the time evolution in the subspace of states
of an unstable particle and its antiparticle need not be equal at
$t > t_{0}$ ($t_{0}$ is the instant of creation of the pair) when
the total system under consideration is CPT invariant but CP
noninvariant. To achieve this we use the transition amplitudes for
transitions $|{\bf 1}> \rightarrow |{\bf 2}>$, $|{\bf 2}>
\rightarrow |{\bf 1}>$ together with the identity expressing the
effective Hamiltonian by these amplitudes and their derivatives
with respect to time $t$. This identity must be fulfilled by any
effective Hamiltonian (both approximate and exact) derived for the
two state complex. The unusual consequence of this result is that,
contrary to the properties of stable particles, the masses of the
unstable particle "1" and its antiparticle "2" need not be equal
for $t \gg t_{0}$ in the case of preserved CPT and violated CP
symmetries.
\end{abstract}
$^{\ast}$e--mail: K.Urbanowski@proton.if.uz.zgora.pl;
K.Urbanowski@if.uz.zgora.pl

\section{Introduction}
The knowledge of all the subtleties of the difference between a
particle and its antiparticle has a fundamental meaning for
understanding our Universe. One can not exclude that the more
complete understanding of this difference can be helpful in
explaining the problem why the observed Universe contains
(according to the physical and astrophysical data) an excess of
matter over antimatter. General properties of antiparticles follow
from properties of particles through the CPT symmetry. The CPT
symmetry is a fundamental theorem (called the CPT Theorem) of
axiomatic quantum field theory, which follows from locality,
Lorentz invariance, and unitarity \cite{cpt}. One consequence of
the CPT theorem is that under the product of operations ${\cal C},
\, {\cal P}$ and $\cal T$ the total Hamiltonian $H$ of the system
considered must be invariant. From this property  one usually
infers that stable particles and the related antiparticles have
exactly the same mass. Indeed, let $H$ be the selfadjoint
Hamiltonian acting in the Hilbert space $\cal H$ of states of the
system under consideration. Assuming that the normalized
eigenstates $|{\psi}_{m}>$ of $H$ for eigenvalues $m$
\begin{equation}
H|{\psi}_{m}> = m |{\psi}_{m}>, \label{psi-m}
\end{equation}
correspond to the particle states, one obtains  vectors
$|{\overline{\psi}}_{\overline{m}}>$ describing the states of the
related antiparticles from vectors $|{\psi}_{m}>$ using the
antiunitary operator $\Theta \stackrel{\rm def}{=} {\cal C}{\cal
P}{\cal T}$ \cite{cpt,messiah+others},
\begin{equation}
\Theta |{\psi}_{m}> = e^{- i \theta}
|{\overline{\psi}}_{\overline{m}}>. \label{anty-psi-m}
\end{equation}
Now, if the relation
\begin{equation}
\Theta H {\Theta}^{-1} = H, \label{[CPT,H]}
\end{equation}
holds, then one infers from (\ref{psi-m}) that every
$|{\overline{\psi}}_{\overline{m}}>$ is also an eigenvector of $H$ for
the same eigenvalue $m$ and thus that the mass $\overline{m}$ of
the antiparticle, whose state is described by
$|{\overline{\psi}}_{\overline{m}}>$, is equal to the mass $m$ of the
related particle  described by vector $|{\psi}_{m}>$,
\begin{equation}
m  \equiv  < {\psi}_{m}|H|{\psi}_{m}>
=
<{\overline{\psi}}_{\overline{m}}|H|{\overline{\psi}}_{\overline{m}}>
\equiv \overline{m}. \label{m=m}
\end{equation}
(We use $\hbar = c = 1$ units).

The properties of singularities of scattering
amplitudes appearing in the S--matrix theory provide one with reasons
for a conclusion similar to (\ref{m=m}). Generally, such a conclusion is
considered to be universal.

The property (\ref{m=m}) of the particle--antiparticle pairs,
obvious for stable particles, is considered to be obligatory also
for unstable particles, though relations of type (\ref{m=m}) can
not be extended to the case of unstable particles because there
are no asymptotic states for these particles and the states of
such particles are described by vectors belonging to $\cal H$,
which are not eigenvectors for the considered Hamiltonian $H$. An
analogous conclusion to the relation (\ref{m=m}) that masses of
unstable particles and their antiparticles are equal follows from
the properties of the Lee, Oehme and Yang (LOY) method of
approximate description
of time evolution in the subspace of states of a two particle subsystem
prepared at some initial instant $t_{0} > - \infty $ and then evolving with
time $t > t_{0}$ \cite{Lee1,Lee2}.

Searching for properties of the unstable
particle--antiparticle pairs one usually uses an effective
nonhermitean Hamiltonian \cite{Lee1} --- \cite{improved}, say
$H_{\parallel}$, which in general can depend on time $t$
\cite{horwitz},
\begin{equation}
H_{\parallel} \equiv M - \frac{i}{2} \Gamma, \label{H||1}
\end{equation}
where
\begin{equation}
M = M^{+}, \; \; \Gamma = {\Gamma}^{+}, \label{M-G}
\end{equation}
are $(2 \times 2)$ matrices, acting in a two--dimensional subspace
${\cal H}_{\parallel}$ of the total state space $\cal H$ and $M$
is called the mass matrix, $\Gamma$ is the decay matrix \cite{Lee1}
--- \cite{improved}. The standard method of derivation of a such
$H_{\parallel}$ bases on a modification of the Weisskopf--Wigner
(WW) approximation \cite{ww}. Lee, Oehme and Yang have adapted the
WW approach to the case of a two particle subsystem
\cite{Lee1,Lee2} to obtain their (approximate) effective
Hamiltonian $H_{\parallel} \equiv H_{LOY}$. Almost all properties
of the neutral kaon complex, or another particle--antiparticle
subsystem can be described by solving the Schr\"{o}dinger--like
evolution equation \cite{Lee1} --- \cite{tsai}
\begin{equation}
i \frac{\partial}{\partial t} |\psi ; t >_{\parallel} =
H_{\parallel} |\psi ; t >_{\parallel}, \; \; (t \geq t_{0} > -\infty),
\label{eq1}
\end{equation}
with the initial condition \cite{tsai,ijmp95}
\begin{equation}
{\parallel} \, |{\psi} ; t = t_{0} >_{\parallel} {\parallel} = 1,
\; \; | {\psi}; t < t_{0} >_{\parallel} = 0, \label{init}
\end{equation}
for $| \psi ; t >_{\parallel}$ belonging to the subspace
${\cal H}_{\parallel} \subset {\cal H}$ spanned, e.g., by
orthonormal neutral  kaons states $|K_{0}>, \; |{\overline{K}}_{0}>$,
and so on, (then states corresponding to the decay products belong
to ${\cal H} \ominus{\cal H}_{\parallel} \stackrel{\rm def}{=}
{\cal H}_{\perp}$).

Solutions of Eq. (\ref{eq1}) can be written in a matrix  form  and
such  a  matrix  defines  the evolution    operator
(which    is     usually     nonunitary) $U_{\parallel}(t)$ acting in
${\cal H}_{\parallel}$:
\begin{equation}
|\psi ; t >_{\parallel} = U_{\parallel}(t)
|\psi ;t_{0} = 0 >_{\parallel} \stackrel{\rm def}{=}
U_{\parallel}(t) |\psi >_{\parallel}, \label{psi||t}
\end{equation}
where,
\begin{equation}
|\psi >_{\parallel} \equiv a_{1}|{\bf 1}> + a_{2}|{\bf 2}>,
\label{init1}
\end{equation}
and $|{\bf 1}>$ stands for vectors of the   $|K_{0}>,  \;
|B_{0}>$, $|n>$ --- neutrons, etc., type and $|{\bf 2}>$
denotes antiparticles  of  the
particle "1": $|{\overline{K}}_{0}>, \; |{\overline{B}}_{0}>
\, |\overline{n} >$, and so
on, $<{\bf j}|{\bf k}> = {\delta}_{jk}$, $j,k =1,2$.

In many papers it
is assumed that the real parts, $\Re (.)$, of the diagonal matrix elements
of $H_{\parallel}$:
\begin{equation}
\Re \, (h_{jj} )
\equiv M_{jj}, \; \;(j =1,2),
\label{m-jj}
\end{equation}
where
\begin{equation}
h_{jk}  =  <{\bf j}|H_{\parallel}|{\bf k}>, \; (j,k=1,2),
\label{h-jk}
\end{equation}
correspond to the masses  of particle "1" and its antiparticle "2"
respectively \cite{Lee1} --- \cite{improved},
(and such an
interpretation of $\Re \, (h_{11})$ and $\Re \, (h_{22})$ will be
used in this paper), whereas the imaginary parts, $\Im (.)$,
\begin{equation}
-2 \Im \, (h_{jj}) \equiv {\Gamma}_{jj}, \; \;(j =1,2),
\label{g-jj}
\end{equation}
are interpreted as the decay widths of these particles \cite{Lee1}
--- \cite{improved}. Such an interpretation seems to be consistent
with the recent and the early experimental data for neutral kaon
and similar complexes \cite{data}.

Taking $H_{||} = H_{LOY}$ and assuming that the property
(\ref{[CPT,H]}) holds in the system considered one easily finds
the standard result of the LOY approach
\begin{equation}
h_{11}^{LOY}  = h_{22}^{LOY}, \label{LOY-h=h}
\end{equation}
which, among others, means that
\begin{equation}
M_{11}^{LOY} = M_{22}^{LOY}, \label{LOY-m=m}
\end{equation}
where $M_{jj}^{LOY} = \Re \, (h_{jj}^{LOY})$ and $h_{jj}^{LOY} =
<{\bf j}|H_{LOY}|{\bf j}>$, ($j =1,2$). This last relation is
interpreted as the equality of masses of the unstable particle
$|{\bf 1}>$ and its antiparticle $|{\bf 2}>$. However, one should
remember that in fact such a conclusion follows from the
approximate expressions for the $H_{LOY}$.
The accuracy of the LOY approximation \cite{improved} is not
sufficient for considering consequences of the relation (\ref{LOY-m=m})
as the universal one.

Note that vectors $|{\psi}_{m}>$ and
$|{\overline{\psi}}_{\overline{m}}>$, (\ref{psi-m}),
(\ref{anty-psi-m}), describe stationary (bound) states of the
system considered. Within this problem the time $t$ can vary from
$t = - \infty$ up to $t = +\infty$. On the other hand, the
following supposition seems to be reasonable: The behaviour of a
system in which time $t$ varies from $t = - \infty$ to $t = +
\infty$, and a system in which $t$ can vary only from $t =t_{0} >
- \infty$ to $t = + \infty$, under ${\cal C}{\cal P}{\cal T}$
transformation need not be the same.

\section{Implications of the Khalfin Theorem}

The aim of this Section is to show that diagonal matrix elements
of the exact effective Hamiltonian $H_{||}$ can not be equal when
the total system under consideration is CPT invariant but CP
noninvariant.

Universal
properties of the (unstable) particle--antiparticle subsystem
of the system described by the  Hamiltonian $H$, for which  the
relation (\ref{[CPT,H]}) holds, can be extracted from the matrix elements of
the exact $U_{||}(t)$ appearing in (\ref{psi||t}). Such $U_{||}(t)$
has the following form
\begin{equation}
U_{||}(t) = P U(t)P, \label{U||}
\end{equation}
where
\begin{equation}
P \stackrel{\rm def}{=} |{\bf 1}><{\bf 1}| + |{\bf 2}><{\bf 2}|,
\label{P}
\end{equation}
and $U(t)$ is
the total unitary evolution operator $U(t)$, which  solves the
Schr\"{o}\-din\-ger equation
\begin{equation}
i \frac{\partial}{\partial t} U(t)|\phi > = H U(t) |\phi >,
\; \; \; U(0) = I,
\label{Schrod}
\end{equation}
$I$ is
the unit operator in $\cal H$, $|\phi > \equiv |\phi ;t_{0} =0> \in
{\cal H}$ is the initial state of the system. Operator $U_{||}(t)$ acts in
the subspace of unstable states ${\cal H}_{||} \equiv P {\cal H}$. Of
course, $U_{||}(t)$ has nontrivial form only if
\begin{equation}
[P, H] \neq 0, \label{[P,H]}
\end{equation}
and only then transitions of states from ${\cal H}_{||}$ into
${\cal H}_{\perp}$ and vice versa, i.e., decay and regeneration processes,
are allowed.

Using the matrix representation one finds
\begin{equation}
U_{||}(t) \equiv \left(
\begin{array}{cc}
{\rm \bf A}(t) & {\rm \bf 0} \\
{\rm \bf 0} & {\rm \bf 0}
\end{array} \right)
\label{A(t)}
\end{equation}
where ${\rm \bf 0}$ denotes the suitable zero submatrices and
a submatrix
${\rm \bf A}(t)$ is the $2 \times 2$ matrix acting in ${\cal H}_{||}$
\begin{equation}
{\rm \bf A}(t) = \left(
\begin{array}{cc}
A_{11}(t) & A_{12}(t) \\
A_{21}(t) & A_{22}(t)
\end{array} \right) \label{A(t)=}
\end{equation}
and $A_{jk}(t) = <{\bf j}|U_{||}(t)|{\bf k}>
\equiv <{\bf j}|U(t)|{\bf k}>$, $(j,k =1,2)$.

Now assuming (\ref{[CPT,H]}) and using, e.g., the following phase
convention \cite{Lee1} --- \cite{LOY},
\begin{equation}
\Theta |{\bf 1}> \stackrel{\rm def}{=} - |{\bf 2}>, \; \; \Theta
|{\bf 2}> \stackrel{\rm def}{=} - |{\bf 1}>, \label{CPT1=2}
\end{equation}
one easily finds that \cite{chiu} --- \cite{nowakowski}
\begin{equation}
A_{11}(t) = A_{22}(t).
\label{A11=A22}
\end{equation}
Note that assumptions (\ref{[CPT,H]}) and (\ref{CPT1=2}) give no
relations between $A_{12}(t)$ and $A_{21}(t)$.

The important relation between amplitudes $A_{12}(t)$ and
$A_{21}(t)$ follows from the famous Khalfin's Theorem \cite{chiu}
--- \cite{leonid2}. This Theorem states that in the case of
unstable states, if amplitudes $A_{12}(t)$ and $A_{21}(t)$ have
the same time dependence
\begin{equation}
r(t) \stackrel{\rm def}{=} \frac{A_{12}(t)}{A_{21}(t)} = {\rm
const} \equiv r, \label{r=const},
\end{equation}
then it must be $|r| = 1$.

For unstable particles the relation (\ref{A11=A22}) means that
decay laws
\begin{equation}
p_{j}(t) \stackrel{\rm def}{=} |A_{jj}(t)|^{2},
\label{p-j}
\end{equation}
(where $j = 1,2$),
of the particle $|{\bf 1}>$ and its antiparticle $|{\bf 2}>$ are equal,
\begin{equation}
p_{1}(t) \equiv p_{2}(t).
\label{p1=p2}
\end{equation}
The consequence of this last  property is that the decay rates of
the particle $|{\bf 1}>$ and its antiparticle $|{\bf 2}>$ must be equal too.

From (\ref{A11=A22}) it does not follow that the masses of the particle "1"
and the antiparticle "2" should be equal. Indeed, every amplitude
$A_{jk}(t)$ is a complex number such that $|A_{jk}(t)| \leq 1$. This means
that, e.g,
\begin{equation}
A_{jj}(t) = e^{{\textstyle -g_{j}(t) - i {\cal M}_{j}(t)}},
\label{g-M}
\end{equation}
where $g_{j}(t), {\cal M}_{j}(t)$ are real functions of $t$, and
$g_{j}(t) \geq 0$. These functions can be connected with the decay
rate, ${\gamma}_{j}(t) \stackrel{\rm def}{=} - \frac{1}{p_{j}(t)}
\frac{\partial p_{j}(t)}{\partial t}$, and mass (energy),
$M_{j}(t)$, of the particle $"j"$ respectively. There is
\begin{equation}
{\gamma}_{j}(t) = 2 \frac{\partial g_{j}(t)}{\partial t},
\label{gamma} \end{equation} which in the case of $g_{j}(t) \equiv
\frac{1}{2} t {\gamma}^{WW}_{j}$ (within the WW approximation, or
the LOY approximation) leads to ${\gamma}_{j}(t) \equiv
{\gamma}_{j}^{WW} = {\rm const}$. In general
\begin{equation}
g_{j}(t) \sim  {\gamma}_{j}(t), \; \; \;
{\cal M}_{j}(t) \sim M_{j}(t), \; \; (t \geq t_{0} =0).
\label{gamma-M}
\end{equation}

From (\ref{A11=A22}) and
(\ref{p1=p2}) it only follows that the decay rates must be equal in the
case considered,
\begin{equation}
{\gamma}_{1}(t) = {\gamma}_{2}(t), \label{g1=g2}
\end{equation}
which need not be true for the
masses. Indeed, one finds that
\begin{equation}
{\cal M}_{1}(t) = {\cal M}_{2}(t) \pm 2n \pi, \; \; (t >0),
\; \; \; (n =0,1,2, \ldots),
\label{M=M+2pi}
\end{equation}
so, it can occur
\begin{equation}
M_{1}(t) =  M_{2}(t), \; \; \; \; {\rm or,} \; \; \; \;
M_{1}(t) \neq  M_{2}(t),
\label{M=M+2pi(a)}
\end{equation}
in a CPT invariant system for $t>0$. Note that in the case of the
unstable particle--antiparticle pair, the property (\ref{A11=A22})
is the only exact relation following from the CPT invariance. So,
in the case of unstable particles connected by a relation
of type (\ref{CPT1=2}), when the property (\ref{[CPT,H]}) holds,
the only exact  relation (\ref{A11=A22}) allows the pair unstable
particle--antiparticle to have different masses.

More conclusions about the properties of the matrix elements of
$H_{||}$, that is in particular about $M_{jj}$, one can infer
analyzing the following identity \cite{horwitz,bull} ---
\cite{pra}
\begin{equation}
H_{||} \equiv H_{||}(t) =
i \frac{\partial U_{||}(t)}{\partial t} [U_{||}(t)]^{-1},
\label{H||2a}
\end{equation}
where $[U_{||}(t)]^{-1}$ is defined as follows
\begin{equation}
U_{||}(t) \, [U_{||}(t)]^{-1} = [U_{||}(t)]^{-1} \, U_{||}(t) \, = \, P.
\label{U^-1}
\end{equation}
(Note that the identity (\ref{H||2a}) holds, independent of
whether $[P,H] \neq 0$ or $[P,H]=0$). The expression (\ref{H||2a})
can be rewritten using the matrix ${\bf A}(t)$

\begin{equation}
H_{||}(t) \equiv  i \frac{\partial {\bf A}(t)}{\partial t}
[{\bf A}(t)]^{-1}. \label{H||2b}
\end{equation}
Relations (\ref{H||2a}), (\ref{H||2b}) must be fulfilled by the
exact as well as by every approximate effective Hamiltonian
governing the time evolution in every two dimensional subspace
${\cal H}_{||}$ of states $\cal H$ \cite{horwitz,bull} ---
\cite{pra}.

It is easy to find from (\ref{H||2a}) the general formulae for the
diagonal matrix elements, $h_{jj}$, of $H_{||}(t)$, in which we are
interested. We have
\begin{eqnarray}
h_{11}(t) &=& \frac{i}{\det {\bf A}(t)} \Big( \frac{\partial
A_{11}(t)}{\partial t} A_{22}(t) - \frac{\partial
A_{12}(t)}{\partial t} A_{21}(t) \Big), \label{h11=} \\
h_{22}(t)
& = & \frac{i}{\det {\bf A}(t)} \Big( - \frac{\partial
A_{21}(t)}{\partial t} A_{12}(t) + \frac{\partial
A_{22}(t)}{\partial t} A_{11}(t) \Big). \label{h22=}
\end{eqnarray}
Now, assuming (\ref{[CPT,H]}) and using the consequence (\ref{A11=A22}) of
this assumption, one finds
\begin{equation}
h_{11}(t) - h_{22}(t) =  \frac{i}{\det {\bf A}(t)} \Big(
\frac{\partial A_{21}(t)}{\partial t} A_{12}(t) - \frac{\partial
A_{12}(t)}{\partial t} A_{21}(t) \Big). \label{h11-h22=}
\end{equation}
Next, after some algebra one obtains
\begin{equation}
h_{11}(t) - h_{22}(t) = - i \, \frac{A_{12}(t) \, A_{21}(t) }{\det
{\bf A}(t)} \; \frac{\partial}{\partial t} \ln
\Big(\frac{A_{12}(t)}{A_{21}(t)} \Big). \label{h11-h22=1}
\end{equation}
This result means that in the considered case for $t>0$
the following Theorem holds:
\begin{equation}
h_{11}(t) - h_{22}(t) = 0 \; \; \Leftrightarrow \; \;
\frac{A_{12}(t)}{A_{21}(t)}\;\; = \; \; {\rm const.}, \; \; (t >
0). \label{h11-h22=0}
\end{equation}
Thus for $t > 0$ the problem under studies is reduced to the
Khalfin's Theorem (see the relation (\ref{r=const})).

From (\ref{h11=}) and (\ref{h22=}) it is easy to see that at $t=0$
\begin{equation}
h_{jj}(0) = <{\bf j}|H|{\bf j}>, \; \; (j=1,2),
\label{hjjt=0}
\end{equation}
which means that
in a CPT invariant system (\ref{[CPT,H]})
in the case of pairs of unstable particles, for which
transformations of type (\ref{CPT1=2}) hold
\begin{equation}
M_{11}(0) = M_{22}(0) \equiv <{\bf 1}|H|{\bf 1}>,
\label{M11=M22}
\end{equation}
the unstable particles "1" and "2" are created at $t=t_{0} \equiv
0$ as  particles with equal masses. The same result can be
obtained from the formula (\ref{h11-h22=1}) by taking $t
\rightarrow 0$. Note that the properties of the function ${\cal
M}_{j}(t)$ following from the definition (\ref{g-M}) and from the
properties of the amplitude $A_{jj}(t)$ and their derivative at
$t=0$ do not contradict these last conclusions.

Now let us go on to analyze the  conclusions
following from the Khalfin's Theorem. CP noninvariance requires
that $|r| \neq 1$ \cite{leonid2,chiu,leonid1,nowakowski}
(see also \cite{Lee1} --- \cite{LOY}, \cite{tsai,data}).
This means that in such a case
it must be $r = r(t) \neq {\rm const.}$.
So, if in the system considered
the property (\ref{[CPT,H]}) holds but
\begin{equation}
[{\cal CP}, H] \neq 0, \label{[CP,H]}
\end{equation}
and the unstable states "1" and "2" are connected by a relation of
type (\ref{CPT1=2}), then at $t > 0$ it must be $(h_{11}(t) -
h_{22}(t)) \neq 0$ in this system. Assuming the LOY interpretation
of $\Re \,(h_{jj}(t))$, ($j=1,2$), one can conclude from the
Khalfin's Theorem and from the property (\ref{h11-h22=0}) that if
$A_{12}(t), A_{21}(t) \neq 0$ for $t > 0$ and if the total system
considered is CPT--invariant, but CP--noninvariant, then
$M_{11}(t) \neq M_{22}(t)$ for $t >0$, that is, that contrary to
the case of stable particles (the bound states), the masses of the
simultaneously created unstable particle "1" and its antiparticle
"2", which are connected by the relation (\ref{CPT1=2}), need not
be equal  for $t>t_{0} =0$.  Of course, such a conclusion
contradicts the standard LOY result (\ref{LOY-h=h}),
(\ref{LOY-m=m}). However, one should remember that the LOY
description of neutral $K$ mesons and similar complexes is only an
approximate one, and that the LOY approximation is not perfect. On
the other hand the relation (\ref{h11-h22=0}) and the Khalfin's
Theorem follow from the basic principles of the quantum theory and
are rigorous. Consequently, their implications should also be
considered rigorous.

\section{Discussion}
In fact there is nothing strange in the above conclusions about
the masses of unstable particles under consideration. From
(\ref{[CPT,H]}) (or from the CPT Theorem \cite{cpt} ) it only
follows that the masses of particle and antiparticle eigenstates
of $H$ (i.e., masses of stationary states for $H$) should be the
same in the CPT invariant system. Such a conclusion can not be
derived from (\ref{[CPT,H]}) for particle $|{\bf 1}>$ and its
antiparticle $|{\bf 2}>$ if they are unstable, i.e., if states
$|{\bf 1}>, |{\bf 2}>$ are not eigenstates of $H$. Simply the
proof of the CPT Theorem makes use of the properties of asymptotic
states \cite{cpt}. Such states do not exist for unstable
particles. What is more, one should remember that the CPT Theorem
of axiomatic quantum field theory has been proved for quantum
fields corresponding to stable quantum objects and only such
fields are considered in the axiomatic quantum field theory. There
is no axiomatic quantum field theory of unstable quantum
particles. So, all implications of the CPT Theorem (including
those obtained within the S--matrix method) need not be valid for
decaying particles prepared at some initial instant $t_{0} = 0$
and then evolving in time $t \geq 0$. Simply, the consequences of
CPT invariance need not be the same for systems in which time $t$
varies from $t = - \infty$ to $t = + \infty$ and for systems in
which $t$ can vary only from $t = t_{0} > - \infty$ to $t = +
\infty$. Similar doubts about the fundamental nature of the CPT
Theorem were formulated in \cite{kobayashi}, where the
applicability of this theorem for QCD was considered. One should
also remember that  conclusions about the equality of masses of stable
particles and their antiparticles following from the properties of
the S-matrix can not be extrapolated to the case of unstable states.
Simply, there is no S--matrix for unstable states.

The following conclusion can be drawn from (\ref{M11=M22}) and
(\ref{h11-h22=0}): In the case when the par\-ti\-cle and
anti\-parti\-cle  states are connected by the relation
(\ref{CPT1=2}) and time $t$ can vary only from $t = t_{0} > =
\infty$ to $t = + \infty$, the time evolution can cause the masses
of the unstable particle and its simultaneously prepared at $t =
0$ antiparticle to be different at $t \gg t_{0}$ in the
CPT--invariant but CP--noninvariant system.
There are other possibilities for interpreting the results
discussed in the present paper. The first is that the
interpretation of the real and imaginary parts of $h_{jj}(t)$ as
the masses and decay rates respectively is wrong. This, however is
highly improbable as the experimental data for neutral $K$-- and
$B$--mesons corroborate the standard interpretation of $\Re
\,(h_{jj})$ and $\Im \, (h_{jj})$. Moreover, such an
interpretation follows directly from (\ref{H||2a}). One can think
of yet another explanation: From the logical and mathematical
point of view quantum mechanics is not selfconsistent. The
possibility of the above statement being true is very low indeed.

The consequences of the exact relation  (\ref{h11-h22=0}) and of
the Khalfin's Theorem confirm the results obtained in  \cite{xx},
\cite{plb,is} within a different method and in \cite{improved},
\cite{pla} --- \cite{ijmpa98}, where a more accurate approximation
(based on the Krolikowski--Rzewuski equation for a distinguished
component of state vector \cite{KR}) than the LOY was used. In
particular, this explains why for the Fridrichs--Lee model
\cite{chiu,ijmpa93} assuming CPT--invariance (i.e.,
(\ref{[CPT,H]}) ) and (\ref{[CP,H]}) it was  found  in
\cite{ijmpa98} that
\begin{eqnarray}
h_{11}^{FL} - h_{22}^{FL} \equiv
\Re \, (h_{11}^{FL}  -  h_{22}^{FL})
& \simeq &
i \frac{ m_{21}{\Gamma}_{12} - m_{12}{\Gamma}_{21} }{4(m_{0} - \mu )}
\nonumber \\
& \equiv & \frac{ {\Im} \, (m_{12}{\Gamma}_{21}) }{2(m_{0}- \mu )},
\label{FL2}
\end{eqnarray}
where $h_{jj}^{FL}, \; (j =1,2)$, denotes $h_{jj}(t \rightarrow \infty)$
calculated for the
Fridrichs--Lee model within  using the more accurate
approximation mentioned above
\cite{ijmpa93,ijmpa95,ijmpa98},
$m_{jk} \equiv H_{jk} = <{\bf j}|H|{\bf k}>$, ($j,k =1,2$),
$m_{0} \equiv H_{11} = H_{22}$ and  $\mu$ denotes the mass of the decay
products. Formulae (\ref{FL2}) were obtained
in \cite{ijmpa98} assuming that
$|m_{12}|\equiv |H_{12}| \ll (m_{0}- \mu ) $.
For the ${K_{0}},
\overline{K_{0}}$--complex
${\Gamma}_{21} \equiv {\Gamma}_{12}^{\ast} \simeq
\Re \,({\Gamma}_{12})
\simeq \frac{1}{2}{\Gamma}_{s} \sim 3,7 \times 10^{-12}$MeV. This
property, the relation (\ref{FL2}) and
the assumption that $m_{0}$
can be considered as the kaon mass, $m_{0} = m_{K}$,
enable us to find the following
estimation of $\Re \, (h_{11}^{FL}  -  h_{22}^{FL})$ for the neutral
K--system \cite{ijmpa98},
\begin{equation}
\Re \, (h_{11}^{FL}  -  h_{22}^{FL})
\sim 9,25 \times 10^{-15} \Im \, (m_{12})
\equiv 9,25 \times 10^{-15} \Im \, (H_{12}) . \label{FL}
\end{equation}

The assumptions leading to (\ref{FL2}) can be used to obtain
\begin{equation}
h_{11}^{FL} + h_{22}^{FL}  \simeq  2m_{0} - i {\Gamma}_{0} -
\frac{i}{2} \frac{ \Re \,(H_{21} {\Gamma}_{12} )}{m_{0} - \mu},
\label{2h-0}
\end{equation}
(where ${\Gamma}_{0} \equiv {\Gamma}_{11} = {\Gamma}_{22}$ ) which
together with
(\ref{FL2}), (\ref{FL}) give
\begin{equation}
\frac{\Re \, (h_{11}^{FL}  -  h_{22}^{FL}) }
{\Re \, (h_{11}^{FL}  +  h_{22}^{FL}) } \simeq
\frac{ \Im \, (H_{12})}{8m_{0}(m_{0} - \mu )} {\Gamma}_{s}
\sim 9,25 \times 10^{-18} \, (\Im \, (H_{12})) \; [{\rm MeV}]^{-1}.
\label{hz-h0}
\end{equation}

This estimation does not contradict the experimental data for
neutral K mesons \cite{data} and that the effect following from
the relation (\ref{h11-h22=0}) is  very small but nonzero.

Note that the relation (\ref{h11-h22=0}) explains also why the
property (\ref{LOY-h=h}) of $H_{LOY}$  takes place in the case of
preserved CPT symmetry. Simply, there is $r = {\rm const.}$ within
the LOY approximation.

Detailed analysis of assumptions leading to the standard form of
the LOY effective Hamiltonian governing the time evolution in a
two state (two particle) subsystem indicates that some
assumptions, which have been used in the LOY treatment of the
problem, and which the WW theory of single line width uses, should
not be directly applied to the case of two, or more, level
subsystems interacting with the rest of the physical system
considered \cite{improved}. Namely, when one considers the single
line width problem in the WW manner it is quite sufficient to
analyze the smallness of matrix elements of the interaction
Hamiltonian, $H^{(1)}$, only (we have $H = H^{(0)} + H^{(1)}$
within the LOY approach). For the multilevel problem, contrary to
the single line problem, such a smallness does not ensure the
suitable smallness of components of the evolution equations
containing these matrix elements. Moreover, there is no necessity
of taking into account the internal dynamics of the subsystem,
which also has an effect on the widths of levels (on the masses of
the particles) in many levels (many particles) subsystems, in such
a case. The observed masses in a two particle subsystem depend on
the interactions of this subsystem with the rest, but they also
depend on the interactions between the particles forming this
subsystem. This means that contrary to the LOY assumption that
$H_{12} = <{\bf 1}|H|{\bf 2}> = 0$, that is $<{\bf 1}|H^{(1)}|{\bf
2}>=0$, one should consider the implications of the assumption
that $<{\bf 1}|H^{(1)}|{\bf 2}>\neq 0$ when one wants to describe
the real properties of two particle subsystems. One finds assuming
(\ref{[CPT,H]}) and using the more accurate approximation than LOY
approximation that
\begin{equation}
h_{11}(t \rightarrow \infty ) - h_{22}(t \rightarrow \infty ) \neq
0, \label{end}
\end{equation}
if $<{\bf 1}|H^{(1)}|{\bf 2}>\neq 0$
\cite{improved,Piskorski,hep-ph-0202253}. The Fridrichs--Lee model
calculations performed within the mentioned approximation confirm
this conclusion
--- see (\ref{FL2}), (\ref{FL}).
On the other hand one can show that the assumption  $<{\bf
1}|H^{(1)}|{\bf 2}>=0$ implies that the more accurate $H_{||}$ is
equal to $H_{LOY}$ \cite{improved,Piskorski,hep-ph-0202253} and
thus that for amplitudes $A_{12}(t), A_{21}(t)$ obtained within
this approximation one finds $r = {\rm const.}$

In a general case one finds within this more accurate
approximation that the parameters describing properties of the
$K_{0}, {\bar{K}}_{0}$ complex  may be written as the sum of the
suitable LOY theory parameter plus a small correction (see
\cite{ijmpa95} formulae (66) --- (86)). Assuming that $H_{12}$ is
suitably small these corrections are found to be proportional to
$H_{12}$. Thus in the case $H_{12} \equiv$ $<{\bf 1}|H^{(1)}|{\bf
2}>=0$ these parameters obtained by means of the mentioned more
accurate approximation coincide with the LOY theory parameters.

The assumption $<{\bf 1}|H^{(1)}|{\bf 2}>=0$ means in the case of
neutral K-- (or B--) complexes, etc., that the first order
$|\Delta S |= 2$ $K_{0} \rightleftharpoons {\bar{K}}_{0}$ and
similar  transitions are forbidden. On the other hand the
assumption $<{\bf 1}|H^{(1)}|{\bf 2}> \neq 0$ leading to the
result (\ref{end}) means that the hypothetical first order
$|\Delta S| = 2$ transitions are allowed in the considered
complexes. So, if measurable deviations from the LOY predictions
for the neutral K-- (or B--) systems will be detected then the
most plausible interpretation of this result will be the existence
of the interactions allowing the first order $|\Delta S| = 2$
transitions \cite{hep-ph-0202253}.

In view of the argument presented in this paper it may be
necessary to reconsider the conclusions contained, e.g., in
\cite{qm-viol,qm-viol1,5th-f,sr-eqp}. There is also a possibility
that the observation on the masses of unstable particles and their
antiparticles (i.e., that they need not be equal from a certain
point in time after their creation) may be one of the causes of
the observed matter--antimatter asymmetry \cite{m-antim}.

\end{document}